\documentclass{PoS-hep}
\usepackage{amsmath}
\usepackage{bm}
\usepackage{epsfig}
\usepackage{graphics}
\usepackage{upgreek}
\usepackage{amsfonts}
\usepackage{amsbsy}
\usepackage{amscd}
\usepackage{bbm}
\usepackage{eufrak}
\usepackage{cite}
\usepackage{times}

\renewenvironment{subequations}{%
\refstepcounter{equation}%
\setcounter{parentequation}{\value{equation}}%
  \setcounter{equation}{0}
  \ignorespaces
}{%
  \setcounter{equation}{\value{parentequation}}%
  \ignorespacesafterend
}
\newcommand{\eeq}{\end{equation}}
\newcommand{\beq}{\begin{equation}}
\newcommand{\ba}{\begin{array}}
\newcommand{\ea}{\end{array}}
\newcommand{\bea}{\begin{eqnarray}}
\newcommand{\eea}{\end{eqnarray}}
\newcommand{\baq}{\begin{eqnarray}}
\newcommand{\eaq}{\end{eqnarray}}

\newcommand{\beqs}{\begin{subequations}}
\newcommand{\eeqs}{\end{subequations}}
\newcommand{\eec}{\end{center}}
\newcommand{\bec}{\begin{center}}
\newcommand{\eem}{\end{matrix}}
\newcommand{\bem}{\begin{matrix}}
\newcommand{\Eref}[1]{Eq.~(\ref{#1})}
\newcommand{\Sref}[1]{Sec.~\ref{#1}}
\newcommand{\Fref}[1]{Fig.~\ref{#1}}
\newcommand{\Tref}[1]{Table~\ref{#1}}
\newcommand{\cref}[1]{Ref.~\cite{#1}}

\newcommand{\etal}{{\it et al.\/}}

\newcommand\eqs[2]{Eqs.~(\ref{#1}) and (\ref{#2})}

\newcommand\eqss[3]{Eqs.~(\ref{#1}), (\ref{#2}) and (\ref{#3})}

\newcommand{\ftn}{\footnotesize}

\newcommand{\TeV}{{\mbox{\rm TeV}}}

\newcommand{\GeV}{{\mbox{\rm GeV}}}
\newcommand{\eV}{{\mbox{\rm eV}}}
\newcommand{\EeV}{{\mbox{\rm EeV}}}
\def\to{\rightarrow}

\def\llgm{\left\lgroup}
\def\rrgm{\right\rgroup}
\def\lf{\left(}
\def\rg{\right)}
\newcommand\vev[1]{\langle {#1} \rangle}
\newcommand\vevi[1]{\langle {#1} \rangle_{\rm I}}
\newcommand\vevii[1]{\left\langle { #1} \right\rangle_{\rm I}}
\newcommand{\Gr}{\ensuremath{\widetilde{G}}}
\newcommand{\Yb}{\ensuremath{Y_{B}}}
\newcommand{\Yg}{\ensuremath{Y_{3/2}}}

\newcommand{\Vhi}{\ensuremath{\widehat V_{\rm IHI}}}
\newcommand{\Hhi}{\ensuremath{\widehat H_{\rm IHI}}}

\newcommand{\Khi}{\ensuremath{K}}

\newcommand{\mP}{\ensuremath{m_{\rm P}}}

\newcommand{\Mgut}{\ensuremath{M_{\rm GUT}}}
\newcommand{\Qef}{\ensuremath{\Lambda_{\rm UV}}}
\newcommand{\Ggut}{\ensuremath{G_{B-L}}}

\newcommand{\lm}{\ensuremath{\lambda_\mu}}

\def\openone{\leavevmode\hbox{\small1\kern-3.8pt\normalsize1}}

\newcommand{\ck}{\ensuremath{c_R}}

\newcommand{\Gsn}{\ensuremath{\what{\Gamma}_{\rm \dph}}}
\newcommand{\GNsn}{\ensuremath{\what{\Gamma}_{\dph\to N_i^cN_i^c}}}
\newcommand{\Ghsn}{\ensuremath{\what{\Gamma}_{\dph\to \hu\hd}}}
\newcommand{\Gysn}{\ensuremath{\what{\Gamma}_{\dph\to XYZ}}}

\newcommand{\msn}{\ensuremath{\what m_{\rm \dph}}}

\newcommand{\hd}{{\ensuremath{H_d}}}
\newcommand{\hu}{{\ensuremath{H_u}}}

\newcommand{\dphi}{\ensuremath{\what{\delta\phi}}}
\newcommand{\dph}{\ensuremath{\delta\phi}}

\newcommand{\ks}{\ensuremath{k_\star}}
\newcommand{\Ns}{\ensuremath{{\what N_\star}}}
\newcommand{\ns}{\ensuremath{n_{\rm s}}}

\newcommand{\as}{\ensuremath{a_{\rm s}}}
\newcommand{\As}{\ensuremath{A_{\rm s}}}

\newcommand{\rcc}{\ensuremath{R}}
\newcommand{\rce}{\ensuremath{\widehat{R}}}
\newcommand{\Ve}{\ensuremath{\widehat{V}}}

\newcommand{\sni}{\ensuremath{N^c_i}}
\newcommand{\ssni}{\ensuremath{\widetilde N_i^c}}

\newcommand{\aS}{\ensuremath{{\rm a}_S}}
\newcommand{\Ald}{\ensuremath{A_\lambda}}
\newcommand{\am}{\ensuremath{{\rm a}_{3/2}}}

\newcommand{\mrh[1]}{\ensuremath{M_{#1N^c}}}
\newcommand{\lrh[1]}{\ensuremath{\lambda_{#1N^c}}}

\newcommand{\mD[1]}{\ensuremath{m_{#1\rm D}}}
\newcommand{\mn[1]}{\ensuremath{m_{#1\nu}}}

\newcommand{\Whi}{\ensuremath{W_{\rm HI}}}
\newcommand{\Wmu}{\ensuremath{W_\mu}}
\newcommand{\Wrhn}{\ensuremath{W_{\rm RHN}}}

\def\ve{\varepsilon}

\def\bbet{{\bar\beta}}
\def\al{{\alpha}}

\def\n{\bar{n}}

\def\th{{\theta}}
\def\thb{{\bar\theta}}
\def\thn{{\theta_{\Phi}}}

\newcommand{\Trh}{\ensuremath{T_{\rm rh}}}
\newcommand{\sg}{\ensuremath{\phi}}
\newcommand{\tfw}{\ensuremath{f_{\rm W}}}

\newcommand{\ld}{\ensuremath{\lambda}}
\newcommand{\ldu}{\ensuremath{\uplambda}}

\newcommand{\kp}{\ensuremath{\kappa}}

\newcommand{\sgx}{\ensuremath{\phi_\star}}
\newcommand{\sgf}{\ensuremath{\phi_{\rm f}}}

\newcommand{\what}{\ensuremath{\widehat}}
\newcommand{\wtilde}{\ensuremath{\widetilde}}

\newcommand{\se}{\ensuremath{\widehat \phi}}
\newcommand{\sex}{\ensuremath{\widehat{\phi}_\star}}

\newcommand{\geu}{\ensuremath{\widehat g}}

\newcommand{\mgr}{\ensuremath{m_{3/2}}}
\newcommand{\mg}{{\ensuremath{M_{1/2}}}}

\def\Kap{K\"{a}hler potential}

\def\Kaa{K\"{a}hler~}
\def\sub{subplanckian}

\def\bcp{{\sc\small Bicep2}/{\it Keck Array}}
\newcommand{\plk}{{\it Planck}}

\newcommand{\diag}{\ensuremath{{\sf diag}}}
\newcommand{\im}{\ensuremath{{\sf Im}}}

\newcommand{\nsu}{\ensuremath{{N_X}}}

\newcommand{\Gbl}{\ensuremath{G_{B-L}}}

\newcommand{\fr}{\ensuremath{f_{R}}}

\newcommand{\fk}{\ensuremath{F_R}}

\newcommand{\kar}{\ensuremath{K_{1}}}
\newcommand{\kbr}{\ensuremath{K_{2}}}

\newcommand{\phc}{\ensuremath{\Phi}}
\newcommand{\phcb}{\ensuremath{\bar\Phi}}
\newcommand\mtta[4]{\mbox{
$\llgm\bem #1 &#2 \cr #3& #4\eem\rrgm$}}

\newcommand{\bdhh}{{\ensuremath{\normalsize I{\kern-2.9pt H}}}}

\title{\boldmath \bfseries Starobinsky-Type $B-L$ Higgs Inflation Leading
Beyond MSSM}

\ShortTitle{Starobinsky-Type $B-L$ Higgs Inflation Leading Beyond
MSSM}

\author{\speaker{C. Pallis}\\
Laboratory of Physics, Faculty of Engineering, \\ Aristotle
University of Thessaloniki, \\ GR-541 24 Thessaloniki, GREECE \\
E-mail: \email{kpallis@gen.auth.gr}}

\abstract{Models of induced-gravity inflation are formulated
within Supergravity employing as inflaton the Higgs field which
leads to a spontaneous breaking of a $U(1)_{B-L}$ symmetry at
$\Mgut=2\cdot10^{16}~\GeV$. We use a renormalizable
superpotential, fixed by a $U(1)$ R symmetry, and logarithmic or
semi-logarithmic \Kap s with integer prefactors which exhibit a
quadratic non-minimal coupling to gravity. We find inflationary
solutions of Starobinsky type in accordance with the observations.
The inflaton mass is predicted to be of the order of
$10^{13}~\GeV$. The model can be nicely linked to MSSM offering an
explanation of the magnitude of the $\mu$ parameter consistently
with phenomenological data. Also it allows for baryogenesis via
non-thermal leptogenesis, provided that the gravitino is heavier
than about $10~\TeV$.
}

\FullConference{Corfu Summer Institute 2022 "School and Workshops on Elementary Particle Physics and Gravity"\\
                 August 28 - September 7 2022\\
                 Corfu, Greece}

\begin{document}

\section{Introduction}

It is well-known \cite{R2r, nIG, rena} that one of the possible
incarnations of Starobisky-type inflation \cite{R2} in
\emph{Supergravity} ({\sf\ftn SUGRA}) can be relied on the
hypothesis of induced gravity \cite{old,igi,lee}. According to
this, inflation is driven in the presence of a non-minimal
coupling among the inflaton field and the Ricci scalar curvature,
$\fr$, such that the reduced Planck mass $\mP$ is determined by a
large (close to Planckian scale $\mP$) \emph{vacuum expectation
value} ({\sf\ftn v.e.v}) of the inflaton at the end of the slow
roll. This is to be contrasted to the case of non-minimal
\cite{sm, univ, nmh} or pole-induced \cite{so} Higgs inflation
where the v.e.v of inflaton is negligible. In this talk we focus
on the implementation of this scenario employing as inflaton a
Higgs field within an ``elementary'' \emph{Grand Unified Theory}
({\sf \ftn GUT}) which extends the gauge symmetry of the
\emph{Standard Model} ({\sf\ftn SM}) by a $U(1)_{B-L}$ factor
\cite{ighi}. In a such case, the unification condition within
\emph{Minimal Supersymmetric SM} ({\sf\ftn MSSM}) may be employed
to uniquely determined the strength of $\fr$ giving rise to an
economical, predictive and well-motivated setting, thereby called
\emph{Induced-gravity Higgs inflation} ({\sf\ftn IHI}) --
cf.~\cref{igsu5}.

Here, we concentrate on the simplest models of IHI introduced in
\cref{ighi} considering exclusively integer prefactors for the
logarithms included in the \Kap s. The particle physics framework
of our presentation is described in Sec.~\ref{fhim} whereas the
engineering of induced-gravity hypothesis is outlined in
\Sref{igsec}. The inflationary part of this context is
investigated in Sec.~\ref{fhi}. Then, in Sec.~\ref{secmu}, we
explain how the MSSM is obtained as a low energy theory and, in
Sec.~\ref{pfhi}, we outline how the observed \emph{baryon
asymmetry of the universe} ({\ftn\sf BAU}) is generated via
\emph{non-thermal leptogenesis} ({\sf\ftn nTL}). Our conclusions
are summarized in Sec.~\ref{con}. Throughout the text, the
subscript of type $,z$ denotes derivation \emph{with respect to}
(w.r.t) the field $z$ and charge conjugation is denoted by a star.
Unless otherwise stated, we use units where $\mP = 2.433\cdot
10^{18}~\GeV$ is taken unity.

%\section{Set-up}\label{fhim}

\section{Particle Physics Embedding}\label{fhim}

We focus on a ``GUT'' based on $\Ggut=G_{\rm SM}\times
U(1)_{B-L}$, where ${G_{\rm SM}}= SU(3)_{\rm C}\times SU(2)_{\rm
L}\times U(1)_{Y}$ is the gauge group of the SM and $B$ and $L$
denote the baryon and lepton number respectively. We below -- see
Secs.~\ref{fhim1} and \ref{fhim2} -- present the basic ingredients
of our proposal.

\subsection{Superpotential}\label{fhim1}

The superpotential of our model naturally splits into four parts:
\beq W=W_{\rm
MSSM}+\Whi+\Wmu+\Wrhn,\>\>\>\mbox{where}\label{Wtot}\eeq
\paragraph{\hspace*{0.6cm} \sf\ftn (a)} $W_{\rm MSSM}$ is the part
of $W$ which contains the usual terms -- except for the $\mu$ term
-- of MSSM, supplemented by Yukawa interactions among the
left-handed leptons ($L_i$) and $\sni$:
\beqs \beq W_{\rm MSSM} = h_{ijD} {d}^c_i {Q}_j \hd + h_{ijU}
{u}^c_i {Q}_j \hu+h_{ijE} {e}^c_i {L}_j \hd+ h_{ijN} \sni L_j \hu.
\label{wmssm}\eeq
Here the $i$th generation $SU(2)_{\rm L}$ doublet left-handed
quark and lepton superfields are denoted by $Q_i$ and $L_i$
respectively, whereas the $SU(2)_{\rm L}$ singlet antiquark
[antilepton] superfields by $u^c_i$ and ${d_i}^c$ [$e^c_i$ and
$\sni$] respectively. The electroweak Higgs superfields which
couple to the up [down] quark superfields are denoted by $\hu$
[$\hd$]. Note that the introduction of three right-handed
neutrinos, $\sni$, is necessary to cancel the $B - L$ gauge
anomaly.

\paragraph{\hspace*{0.6cm} \sf\ftn (b)} $\Whi$ is the part of $W$ which is relevant for
IHI and takes the form
\beq\label{Whi} \Whi= \ld S\lf \bar\Phi\Phi-M^2/4\rg.\eeq
The imposed $U(1)_R$ symmetry ensures the linearity of $\Whi$
w.r.t $S$. This fact allows us to isolate easily via its
derivative the contribution of the inflaton into the F-term SUGRA
potential, placing $S$ at the origin -- see \Sref{fhi1}. The
inflaton is contained in the system $\bar\Phi - \Phi$. We are
obliged to restrict ourselves to \sub\ values of $\bar\Phi\Phi$
since the imposed symmetries do not forbid non-renormalizable
terms of the form $(\bar\Phi\Phi)^{p}$ with $p>1$ -- see
\Sref{fhi2}.

\paragraph{\hspace*{0.6cm} \sf\ftn (c)} $\Wmu$ is the part of $W$ which is responsible for the generation of the
$\mu$ term of MSSM and takes the form
\beq\label{Wmu} \Wmu=\lm S\hu\hd. \eeq
As $\Whi$, $\Wmu$ is also linear to $S$ and so, the imposed
$U(1)_R$ plays also a key role in the resolution of the $\mu$
problem of MSSM  -- see \Sref{secmu}.

\paragraph{\hspace*{0.6cm} \sf\ftn (d)} $\Wrhn$ is the part of $W$ which provides Majorana
masses for netrinos and reads
\beq\label{Wrhn} \Wrhn=\lrh[i]\phcb N^{c2}_i\,. \eeq\eeqs
The same term assures the decay of the inflaton to $\ssni$, whose
subsequent decay can activate nTL \cite{ntl}. Here, we work in the
so-called \emph{$\sni$-basis}, where $\mrh[i]$ is diagonal, real
and positive. These masses, together with the Dirac neutrino
masses of the forth term in Eq.~(\ref{wmssm}), lead to the light
neutrino masses via the seesaw mechanism -- see \Sref{lept1}.

\renewcommand{\arraystretch}{1.1}

\begin{table}[!t]
\begin{center}
\begin{tabular}{|c|c|c|c|c|}\hline
{\sc Superfields}&{\sc Representations}&\multicolumn{3}{|c|}{\sc
Global Symmetries}\\\cline{3-5}
&{\sc under $\Gbl$}& {\hspace*{0.3cm} $R$\hspace*{0.3cm} }
&{\hspace*{0.3cm}$B$\hspace*{0.3cm}}&{$L$} \\\hline\hline
\multicolumn{5}{|c|}{\sc Matter Fields}\\\hline
{$e^c_i$} &{$({\bf 1, 1}, 1, 1)$}& $1$&$0$ & $-1$ \\
{$N^c_i$} &{$({\bf 1, 1}, 0, 1)$}& $1$ &$0$ & $-1$
 \\
{$L_i$} & {$({\bf 1, 2}, -1/2, -1)$} &$1$&{$0$}&{$1$}
\\
{$u^c_i$} &{$({\bf 3, 1}, -2/3, -1/3)$}& $1$  &$-1/3$& $0$
\\
{$d^c_i$} &{$({\bf 3, 1}, 1/3, -1/3)$}& $1$ &$-1/3$& $0$
 \\
{$Q_i$} & {$({\bf \bar 3, 2}, 1/6 ,1/3)$} &$1$ &$1/3$&{$0$}
\\ \hline
\multicolumn{5}{|c|}{\sc Higgs Fields}\\\hline
{$\hd$}&$({\bf 1, 2}, -1/2, 0)$& {$0$}&{$0$}&{$0$}\\
{$\hu$} &{$({\bf 1, 2}, 1/2, 0)$}& {$0$} & {$0$}&{$0$}\\
\hline
{$S$} & {$({\bf 1, 1}, 0, 0)$}&$2$ &$0$&$0$  \\
{$\Phi$} &{$({\bf 1, 1}, 0, 2)$}&{$0$} & {$0$}&{$-2$}\\
{$\bar \Phi$}&$({\bf 1, 1}, 0,-2)$&{$0$}&{$0$}&{$2$}\\
\hline\end{tabular}
\end{center}
\caption[]{\sl \small Representations under $\Gbl$ and extra
global charges of the superfields of our model.}\label{tab1}
\end{table}
\renewcommand{\arraystretch}{1.}

\subsection{\Kaa\ Potentials}\label{fhim2}

The objectives of our model are feasible if $W$ in \Eref{Wtot}
cooperates with \emph{one} of the following \Kap s:
\beqs\bea
K_1&=&-3\ln\left(\ck(\fk+\fk^*)-\frac{|\phc|^2+|\phcb|^2}{3}+F_{1X}(|X|^2)\right)~~~\mbox{with}~~~F_{1X}=-\ln\left(1+|X|^2/3\right),\label{K1}\\
K_2&=&-2\ln\left(\ck(\fk+\fk^*)-\frac{|\phc|^2+|\phcb|^2}{2}\right)+F_{2X}(|X|^2)~~~\mbox{with}~~~F_{2X}=N_X\ln\left(1+|X|^2/N_X\right),
\label{K2} \eea\eeqs
where $\fk=\Phi\bar\Phi$, $0<N_X<6$, $X^\gamma=S,\hu,\hd,\ssni$
and the complex scalar components of the superfields $\Phi,
\bar\Phi, S, \hu$ and $\hd$ are denoted by the same symbol whereas
this of $\sni$ by $\ssni$. We assume that $X^\gamma$ have
identical kinetic terms expressed by the functions $F_{lX}$ with
$l=1,2$. These functions ensures the stability and the heaviness
of these modes \cite{su11} employing \emph{exclusively} quadratic
terms. Both $K$'s reduce to the same $K_0$ for $X^\al=0$ with the
aid of the frame function $\Omega$ defined as
\beq\label{Ndef} K_0=-N\ln\lf-\frac{\Omega}{N}\rg~~~\mbox{with}
~~~\frac{\Omega}{N}= -\ck(\fk+\fk^*)+\frac{|\phc|^2+|\phcb|^2}{N}
~~~\mbox{and}
~~~N=\begin{cases}3~~~\mbox{for}~~~K=\kar,\\2~~~\mbox{for}~~~K=\kbr.\end{cases}\eeq
Henceforth, $N$ assists us to unify somehow the two $K$'s
considered in Eqs.~(\ref{K1}) and (\ref{K2}).

\section{SUGRA Version of Induced-Gravity Conjecture}\label{igsec}

The scale $M$ and the function $\fk$ involved in
\eqss{Whi}{K1}{K2} assist us in the implementation of the idea of
induced gravity. To explain how it works, we introduce our
notation in the two relevant frames in \Sref{ig1} and then, in
\Sref{ig2}, we derive the SUSY vacuum which plays a key role
imposing the induced-gravity condition -- see \Sref{ig3}.

\subsection{From Einstein to Jordan Frame}\label{ig1}

We concentrate on $\Whi$ and extract the part of the
\emph{Einstein frame} ({\sf\ftn EF}) action within SUGRA related
to the complex scalars $z^\al=S,\phc,\phcb$. This has the form
\cite{ighi}
\beq\label{Saction1}  {\sf S}=\int d^4x \sqrt{-\what{
\mathfrak{g}}}\lf-\frac{1}{2}\rce +K_{\al\bbet} \geu^{\mu\nu}D_\mu
z^\al D_\nu z^{*\bbet}-\Ve_{\rm SUGRA}\rg\,, \eeq
where $\rce$ is the EF Ricci scalar curvature, $D_\mu$ is the
gauge covariant derivative, $K_{\al\bbet}=K_{,z^\al z^{*\bbet}}$,
and $K^{\al\bbet}K_{\bbet\gamma}=\delta^\al_{\gamma}$ and
$\mathfrak{g}$ is the determinant of the EF metric
$\geu^{\mu\nu}$. Also, $\Ve$ is the EF SUGRA potential which can
be found in terms of $\Whi$ in \Eref{Whi} and the $K$'s in
Eqs.~(\ref{K1}) -- (\ref{K2}) via the formula
\beqs\beq \Ve_{\rm SUGRA}=\Ve_{\rm F}+\Ve_{\rm
D}~~~\mbox{with}~~~\Ve_{\rm F}=e^{\Khi}\left(K^{\al\bbet}(D_\al
W_{\rm HI})D^*_\bbet W_{\rm HI}^*-3{\vert W_{\rm
HI}\vert^2}\right)~~~\mbox{and}~~~\Ve_{\rm D}=\frac{g_{BL}^2}2
{\rm D}_{BL}^2.\label{Vsugra} \eeq
Here the \Kaa covariant derivative reads $D_\al W_{\rm HI}=W_{{\rm
HI},z^\al}+K_{,z^\al}W_{\rm HI}$ whereas the D term due to $B-L$
symmetry is found to be
\beq\label{dterm} {\rm D}_{BL}=
\lf{|\phc|^2-|\phcb|^2}\rg/({-\Omega/N}).\eeq\eeqs
As induced by \eqs{Ndef}{dterm}, the field configuration
\beq
\vevi{\Phi}=\vevi{\bar\Phi}~~~\mbox{and}~~~\vevi{X^\al}=0,\label{inftr1}\eeq
assures $\vevi{\Ve_{\rm D}}=0$ where the symbol $\vevi{Q}$ denotes
values of a quantity $Q$ along the path of \Eref{inftr1}.
Henceforth, we confine ourselves to this path -- assuming in
addition that $\arg(\phc)=\arg(\phcb)$ -- which is a honest
inflationary trajectory, supporting IHI driven exclusively by
$\Ve_{\rm F}$.

The performance of a conformal transformation after defining the
\emph{Jordan Frame} ({\sf\ftn JF}) metric as
\beq g^{\mu\nu} = -\frac{\Omega}{N} \geu^{\mu\nu}~~~\mbox{yields
\cite{ighi} via \Eref{Saction1}} ~~~~{\sf S}=\int d^4x
\sqrt{-\mathfrak{g}}\lf\frac{
\Omega}{2N}\rcc-\cdots\rg\label{Saction}\eeq
which reveals that $-\Omega/N$ plays the role of a (dimensionless)
non-minimal coupling to gravity -- here we use unhatted symbols
for the JF quantities and the ellipsis includes terms irrelevant
for our discussion. Comparing \Eref{Ndef} with the $K$'s in
Eqs.~(\ref{K1}) and (\ref{K2}) we can infer that the emergence of
Einstein gravity at the vacuum dictates
\beq
-\vev{{\Omega}/{N}}=2(N\ck+1)\vev{\phc}^2/{N}=1,\label{minK1}\eeq
where we assume that $\vev{\phc}$ is included in the inflationary
trough of \Eref{inftr1}. Its value as a function of the model
parameters is calculated in the next section.

\subsection{SUSY Potential}\label{ig2}

The implementation of the IHI requires the generation of $\mP$ at
the vacuum of the theory. It can be determined expanding $V_{\rm
SUGRA}$ in powers of $1/\mP$. Namely, we obtain the following
low-energy effective potential which plays the role of SUSY one
\beqs \beq \label{Vsusy} V_{\rm SUSY}=\vevii{\widetilde
K^{\al\bbet} W_{\rm HI\al} W^*_{\rm HI\bbet}}+\cdots,\eeq
where the ellipsis represents terms proportional to $\Whi$ or
$|\Whi|^2$ which obviously vanish along the path in \Eref{inftr1}.
Also, $\widetilde K$ is the limit of the $K$'s in Eqs.~(\ref{K1})
and (\ref{K2}) for $\mP\to\infty$. The absence of unity in the
arguments of the logarithms multiplied by $N$ in these $K$'s
prevents the drastic simplification of $\wtilde K$ -- cf.
\cref{univ}. As a consequence, the expression of the resulting
$V_{\rm SUSY}$ is rather lengthy. For this reason we confine
ourselves below to $K=K_2$ where $F_{2S}$ is placed outside the
first logarithm in \Eref{K1} and so $\wtilde K$ can be somehow
simplified.  Namely, we get
\beq \label{Kquad}\widetilde K=-N\ln\lf-\Omega/N\rg
+|S|^2\,,\eeq\eeqs
from which we can then compute
\beqs\beq \lf \vevi{\widetilde K_{\al\bbet}}\rg=\diag\lf
\widetilde M_{\phcb\phc},1\rg ~~~\mbox{with}~~~~\widetilde
M_{\phcb\phc}=\frac{2}{\vevi{\Omega}^2}\mtta{(4\ck-1)|\phc|^2}{|2\ck\phc-\phc^*|^2}{|2\ck\phc-\phc^*|^2}{(4\ck-1)|\phc|^2}.
\label{wKab}\eeq
To compute $V_{\rm SUSY}$ we need to know
\beq\vevi{\wtilde K^{\al\bbet}}=\diag\lf \widetilde
M^{-1}_{\phcb\phc},1\rg,~~~\mbox{where}~~~ \widetilde
M_{\phcb\phc}^{-1}=-\frac{\vevi{\Omega}^2}{2{\det\widetilde
M_{\phcb\phc}}}\mtta{-(4\ck-1)|\phc|^2}{|2\ck\phc-\phc^*|^2}{|2\ck\phc-\phc^*|^2}{-(4\ck-1)|\phc|^2},\label{wkin}\eeq
where the prefactor can be explicitly written as
\beq  \frac{\vevi{\Omega}^2}{{\det\widetilde
M_{\phcb\phc}}}=\frac{|\phc|^2-\ck(\phc^2-\phc^{*2})}{\ck(\phc^2+\phc^{*2}-4\ck|\phc|^2)}
\label{detMpm} \eeq\eeqs
Upon substitution of \Eref{wkin} into \Eref{Vsusy} we obtain
\beq V_{\rm SUSY}\simeq\ld^2\left|\phcb\phc-\frac14{M^2}\right|^2+
\frac{\vevi{\Omega}^2}{\det\widetilde
M_{\phcb\phc}}\ld^2|S|^2|\phc|^2\lf(4\ck^2-1)|\phc|^2-|\phc-2\ck\phc^*|^2\rg.
\label{VF}\eeq
We remark that the SUSY vacuum lies along the direction in
\Eref{inftr1} with
\beq \vev{S}=0 \>\>\>\mbox{and}\>\>\>
|\vev{\Phi}|=|\vev{\bar\Phi}|=M/2,\label{vevs} \eeq
where $\vev{S}$ may slightly deviate from its value above after
inclusion of soft SUSY breaking effects -- see \Sref{secmu1}. The
result in \Eref{vevs} holds also for $K=\kar$ as we can verify
after a more tedious computation. From \Eref{vevs} it is clear
that $\vev{\Phi}$ and $\vev{\bar\Phi}$ spontaneously break
$U(1)_{B-L}$ down to $\mathbb{Z}^{B-L}_2$. Note that $U(1)_{B-L}$
is already broken during IHI and so no cosmic string are formed --
see \Sref{fhi2}.

\subsection{Induced-Gravity Requirement}\label{ig3}

Inserting \Eref{vevs} into \Eref{minK1} we deduce that the
conventional Einstein gravity can be recovered at the vacuum if
\beq M=\sqrt{{2N}/{(N\ck-1)}}. \label{ig}\eeq
As we show in \Sref{fhi3}, the GUT requirement offers the
prediction $\ck\sim10^4$. Therefore, the resulting $M$ has a size
comparable to $\mP$ as expected from the establishment of the
theory in \Sref{fhim1}.

\section{Inflationary Scenario}\label{fhi}

The salient features of our inflationary scenario are studied at
tree level in \Sref{fhi1} and at one-loop level in \Sref{fhi2}. We
then present its predictions in \Sref{fhi3}.

\subsection{Inflationary Potential}\label{fhi1}

If we express $\Phi, \bar\Phi$ and $X^\gamma= S,\hu,\hd,\ssni$
according to the parametrization
\beq\label{hpar} \Phi=\sg\, e^{i\th}\cos\thn/\sqrt{2},~~~\bar\Phi=
\sg\,e^{i\thb}\sin\thn/\sqrt{2}~~~\mbox{and}~~~X^\gamma=
\lf{x^\gamma +i\bar x^\gamma}\rg/{\sqrt{2}}\,,~~\mbox{where}~~~
0\leq\thn\leq\pi/2, \eeq
the D-flat direction in \Eref{inftr1} is now expressed as
\beq \label{inftr} x^\gamma=\bar
x^\gamma=\th=\thb=\hu=\hd=\ssni=0\>\>\>\mbox{and}\>\>\>\thn={\pi/4}\,.\eeq
Along this, the only surviving term of $\Ve_{\rm SUGRA}$ in
\Eref{Vsugra} -- after replacing $\Whi$ with $\Whi+\Wmu+\Wrhn$ --
can be written as
\beq \label{Vhi} \Vhi= e^{K}K^{SS^*}\, |W_{{\rm HI},S}|^2=
\frac{\ld^2(\sg^2-M^2)^2}{16\fr^{N}}\cdot\begin{cases}
\fr&\mbox{for}\>\> K= K_1,\\
1&\mbox{for}\>\> K=K_2,\end{cases}
~~~\mbox{where}~~~\fr=-\vevii{\frac{\Omega}{N}}=\frac{(N\ck-1)\sg^2}{2N}\eeq
-- see \Eref{Ndef}. Clearly $\Vhi$ develops an inflationary
plateau as in the original Starobinsky inflationary model
\cite{R2r,plin}. To specify the EF canonically normalized fields,
we note that, for the $K$'s in Eqs.~(\ref{K1}) and (\ref{K2}),
$K_{\al\bbet}$ along the configuration in \Eref{inftr} takes the
form
\beq \label{Kab} \vevi{K_{\al\bbet}}=\diag\lf
M_{\phcb\phc},\underbrace{K_{\gamma\bar\gamma},...,K_{\gamma\bar\gamma}}_{8~\mbox{\ftn
elements}}\rg~~~\mbox{with}~~~
M_{\phcb\phc}=\mtta{\kappa}{\bar\kappa}{\bar\kappa}{\kappa}
~~~\mbox{and}~~~K_{\gamma\bar\gamma}=\begin{cases}
\fr^{-1}&\mbox{for}\>\>\>K=K_1\,,\\
1&\mbox{for}\>\>\>K=K_{2}\,,
\end{cases}\eeq
where $\kp=(1+N\ck)/2\fr$ and $\bar\kp=N/\sg^2$. Upon
diagonalization of $M_{\phcb\phc}$ we find its eigenvalues which
are
\beq \label{kpm}\kp_+=N\ck/\fr \>\>\>\mbox{and}\>\>\>
\kp_-=1/{\fr}.\eeq
Note that the existence of the real terms $|\Phi|^2+|\bar\Phi|^2$
in Eqs.~(\ref{K1}) and (\ref{K2}) is vital for our models, since
otherwise the off diagonal elements of $M_{\phcb\phc}$ would have
been zero, one of the eigenvalues above would have been zero and
so no $M_{\phcb\phc}^{-1}$ could have been defined.

Inserting \eqs{hpar}{Kab} into the kinetic term of ${\sf S}$ in
\Eref{Saction1} we can specify the canonically normalized (hatted)
fields, as follows
\beq \label{VJe} \frac{d\se}{d\sg}=J,~~~\widehat{\theta}_+
={J\over\sqrt{2}}\sg\theta_+,~~\widehat{\theta}_-
=\sqrt{\frac{\kp_-}{2}}\sg\theta_-,~~\widehat \theta_\Phi =
\sqrt{\kp_-}\sg\lf\theta_\Phi-\frac{\pi}{4}\rg~~~\mbox{and}~~~(\what{x}^{\gamma},\what{\bar
x}^{\gamma})=\sqrt{K_{\gamma\bar\gamma}}(x^\gamma,\bar
x^\gamma)\,,\eeq %%
where $J=\sqrt{\kp_+}$ and
$\th_{\pm}=\lf\bar\th\pm\th\rg/\sqrt{2}$. As we show below, the
masses of the scalars besides $\se$ during IHI are heavy enough
such that the dependence of the hatted fields on $\sg$ does not
influence their dynamics.

\subsection{Stability and one-Loop Radiative Corrections}\label{fhi2}

\renewcommand{\arraystretch}{1.2}
 \begin{table}[!t] {\small\bec\begin{tabular}{|c|c|c|c|c|}\hline
{\sc Fields}&{\sc Einge-} & \multicolumn{3}{c|}{\sc Masses
Squared}\\ \cline{3-5} &{\sc states}&& {$K=K_1$} &{$K=K_{2}$}
\\\hline\hline
14 Real &$\widehat\theta_{+}$&$\widehat m_{\theta+}^2$& $4\Hhi^2$
&{$6\Hhi^2$}\\\cline{3-5}
Scalars&$\widehat \theta_\Phi$ &$\widehat m_{ \theta_\Phi}^2$&
{$M^2_{BL}$}&{$M^2_{BL}$}\\\cline{3-5}
&$\widehat s, \widehat{\bar{s}}$ &$ \widehat
m_{s}^2$&$\Hhi^2(\ck\phi^2-9)$&{$6\Hhi^2/N_X$}\\\cline{3-5}
& $\widehat{h}_{\pm},\widehat{\bar h}_{\pm}$ &  $ \widehat
m_{h\pm}^2$&$3\Hhi^2\ck\lf
\sg^2/{6}\pm{2\lm/\ld}\rg$&{$3\Hhi^2\lf1+1/N_X\pm{4\lm}/{\ld\sg^2}\rg$}\\\cline{3-5}
& $\widehat{\tilde\nu}^c_{i}, \widehat{\bar{\tilde\nu}}^c_{i}$ &
$\widehat m_{i\tilde \nu^c}^2$&$3\Hhi^2\ck\lf
\sg^2/6+8\ld^2_{iN^c}/{\ld^2}\rg$&{$3\Hhi^2\lf1+1/N_X+16\ld^2_{iN^c
}/\ld^2\sg^2\rg$}\\\hline
1 Gauge Boson& $A_{BL}$ &$
M_{BL}^2$&\multicolumn{2}{c|}{$2Ng^2/\lf N\ck-1\rg$}\\\hline
$7$ Weyl & $\what{\psi}_\pm$ & $\what m^2_{\psi\pm}$ &
\multicolumn{2}{c|}{$12\Hhi^2/\ck^2\sg^4$}
\\\cline{3-5} Spinors &$\ldu_{BL}, \widehat\psi_{\Phi-}$&
$M_{BL}^2$&\multicolumn{2}{c|}{$2Ng^2/\lf N\ck-1\rg$}\\\cline{3-5}
&${\widehat N_i^c}$& $ \widehat
m_{{iN^c}}^2$&\multicolumn{2}{c|}{$48\Hhi^2\ck\ld^2_{iN^c
}/\ld^2\sg^2$}\\\hline
\end{tabular}\eec}
\hfill \caption{\sl\small The mass squared spectrum of our models
along the path in Eq.~(4.2) for $\sg\ll1$ and $N$'s defined in
Eq.~(2.4).}\label{tab3}
\end{table}
\renewcommand{\arraystretch}{1.}

We can verify that the inflationary direction in \Eref{inftr} is
stable w.r.t the fluctuations of the non-inflaton fields. To this
end, we construct the mass-squared spectrum of the scalars taking
into account the canonical normalization of the various fields in
\Eref{VJe}. In the limit $\ck\gg1$, we find the expressions of the
masses squared $\what m^2_{z^\al}$ (with
$z^\al=\theta_+,\theta_\Phi,x^\gamma$ and $\bar x^\gamma$)
arranged in \Tref{tab3}. These results approach rather well for
$\sg=\sgx$ -- see \Sref{fhi2} -- the quite lengthy, exact
expressions taken into account in our numerical computation. The
various unspecified there eigenstates are defined as follows
\beqs\beq \widehat h_\pm=(\widehat h_u\pm{\widehat
h_d})/\sqrt{2},~~~ \widehat{\bar h}_\pm=(\widehat{\bar
h}_u\pm\widehat{\bar h}_d)/\sqrt{2}~~~\mbox{and}~~~\what \psi_\pm
=(\what{\psi}_{\Phi+}\pm \what{\psi}_{S})/\sqrt{2}, \eeq
where the (unhatted) spinors $\psi_\Phi$ and $\psi_{\bar\Phi}$
associated with the superfields $\Phi$ and $\bar\Phi$ are related
to the normalized (hatted) ones in \Tref{tab3} as follows
\beq \label{psis}
\what\psi_{\Phi\pm}=\sqrt{\kp_\pm}\psi_{\Phi\pm}~~~
\mbox{with}~~~\psi_{\Phi\pm}=(\psi_\Phi\pm\psi_{\bar\Phi})/\sqrt{2}\,.
\eeq\eeqs

From \Tref{tab3} it is evident that $0<\nsu\leq6$ assists us to
achieve $m^2_{{s}}>\Hhi^2=\Vhi/3$ -- in accordance with the
results of \cref{su11} -- and also enhances the ratios
$m^2_{X^{\tilde\gamma}}/\Hhi^2$ for
$X^{\tilde\gamma}=h_+,{\tilde\nu}^c_{i}$ w.r.t the values that we
would have obtained, if we had used just canonical terms in the
$K$'s. On the other hand, $\what m^2_{h-}>0$ implies
\beq\label{lmb} \lm\lesssim\ld\sg^2/4N
~~~\mbox{for}~~~K=\kar~~~\mbox{and}~~~\lm\lesssim\ld\sg^2(1+1/\nsu)/4~~~\mbox{for}~~~K=\kbr\,.\eeq
In both cases, the quantity in the right-hand side of the
inequalities takes its minimal value at $\sg=\sgf$ -- see
\Sref{fhi2} -- and numerically equals to $2\cdot10^{-5}-5\cdot
10^{-6}$. In \Tref{tab3} we display also the mass $M_{BL}$ of the
gauge boson $A_{BL}$ which becomes massive having `eaten' the
Goldstone boson $\th_-$. This signals the fact that $\Ggut$ is
broken during IHI and so no cosmological defects are produced.
Also, we can verify \cite{ighi} that radiative corrections \'a la
Coleman-Weinberg can be kept under control provided that we
conveniently select the relevant renormalization mass scale
involved.

\subsection{SUSY Gauge Coupling Unification}\label{fhi3}

The value of $M_{BL}$ in \Tref{tab3} computed at the vacuum of
\Eref{vevs}, $\vev{M_{BL}}$, may in principle, be unconstrained
since $U(1)_{B-L}$ does not disturb the unification of the MSSM
gauge coupling constants. To be more specific, though, we prefer
to determine $M_{BL}$ by requiring that it takes the value $\Mgut$
dictated by this unification at the vacuum of \Eref{vevs}. Namely,
we impose
\beq \label{Mgut1}
\vev{M_{BL}}=\Mgut\simeq2/2.43\cdot10^{-2}=8.22\cdot10^{-3}\,.\eeq
This simple principle has an important consequence for IHI, since
it implies via the findings of \Tref{tab3}
\beq \ck=\frac1N+\frac{2g_{BL}^2}{\Mgut^2}\simeq1.451\cdot10^4\,,
\label{Mgutr}\eeq
leading to $M\simeq0.0117$ via \Eref{ig}. Here we take
$g_{BL}\simeq0.7$ which is the value of the unified coupling
constant within MSSM.

Although $\ck$ above is very large, there is no problem with the
validity of the effective theory, in accordance with the results
of earlier works \cite{R2r, nIG, lee}. To clarify further this
point, we have to identify the ultraviolet cut-off scale $\Qef$ of
theory analyzing the small-field behavior of our models. Indeed,
expanding about $\vev{\phi}=M$ -- see \Eref{ig} -- the second term
in the r.h.s of \Eref{Saction1} for $\mu=\nu=0$ and $\Vhi$ in
\Eref{Vhi} we obtain
%\begin{align}
\beqs\beq\label{Jexpr}  J^2 \dot\phi^2\simeq\lf 1
-\sqrt{\frac{2}{N}}\dphi+\frac{3}{2N}\dphi^2-\sqrt{\frac{2}{N^3}}\dphi^3
+\cdots\rg\dot{\what{\delta\sg}}^2,\eeq where $\dphi$ is the
canonically normalized inflaton at the vacuum -- see \Sref{lept0}
-- and \beq\Vhi\simeq\frac{\ld^2\dphi^2}{2N\ck^{2}}
\lf1-\frac{2N-1}{\sqrt{2N}}\dphi+\frac{8N^2-4N+1}{8N}\what{\delta\sg}^2+\cdots
\rg. \label{Vexpr}\eeq\eeqs
%\end{align}8N^2-4N+1=4N(2N-1)+1
These expressions indicate that $\Qef=\mP$, since $\ck$ does not
appear in any of their numerators.

\subsection{Inflationary Observables}\label{fhi4}

A period of slow-roll IHI is controlled by the strength of the
slow-roll parameters
\beq\label{sr}\widehat\epsilon= {1\over2}\left(\frac{\Ve_{{\rm
IHI},\se}}{\Vhi}\right)^2\simeq
16\frac{\tfw^2}{N\ck^4\sg^8}~~~~\mbox{and}~~~~\widehat\eta =
\frac{\Ve_{{\rm IHI},\se\se}}{\Vhi} \simeq8\frac{2-\tfw}{N
\tfw^{2}}\,\>\>\>\mbox{with}\>\>\>\tfw=\ck\sg^2-2.\eeq
Expanding $\widehat\epsilon$ and $\widehat\eta$ for $\sg\ll 1$ we
can find that IHI terminates for $\sg=\sgf$ such that
\beq{\ftn\sf max}\{\widehat\epsilon(\sgf),|\widehat\eta(\sgf)|\}=1
~~~\Rightarrow~~~\sg_{\rm f}\simeq \mbox{\sf\small max}
\lf\frac2{\sqrt{\ck\sqrt{N}}},
2\sqrt{\frac{2}{N\ck}}\rg.\label{sgfr}\eeq

The number of e-foldings, $\Ns$, that the pivot scale
$\ks=0.05/{\rm Mpc}$ suffers during IHI can be calculated through
the relation
\begin{equation}
\label{Nhi}  \Ns=\:\int_{\se_{\rm f}}^{\se_\star}\, d\se\:
\frac{\Ve_{\rm IHI}}{\Ve_{\rm IHI,\se}}\simeq
\frac{N\ck}{8}\sgx^2~\Rightarrow~\sgx\simeq
2\lf\frac{2\Ns}{N\ck}\rg^{1/2}\simeq\begin{cases}0.11,&K=K_1,\\0.13,&K=K_2,\end{cases}
\end{equation}
where $\sex$ [$\sgx$] with $\sgx\gg\sgf$ is the value of $\se$
[$\sg$] when $\ks$ crosses the inflationary horizon. Thanks to
large $\ck$ in \Eref{Mgutr}, $\sgx\ll1$ and therefore, our
proposal is automatically well stabilized against corrections from
higher order terms of the form $(\phc\phcb)^p$ with $p>1$ in
$\Whi$ -- see \Eref{Whi}.

\begin{figure}[!t]\vspace*{-.12in}
\bec\epsfig{file=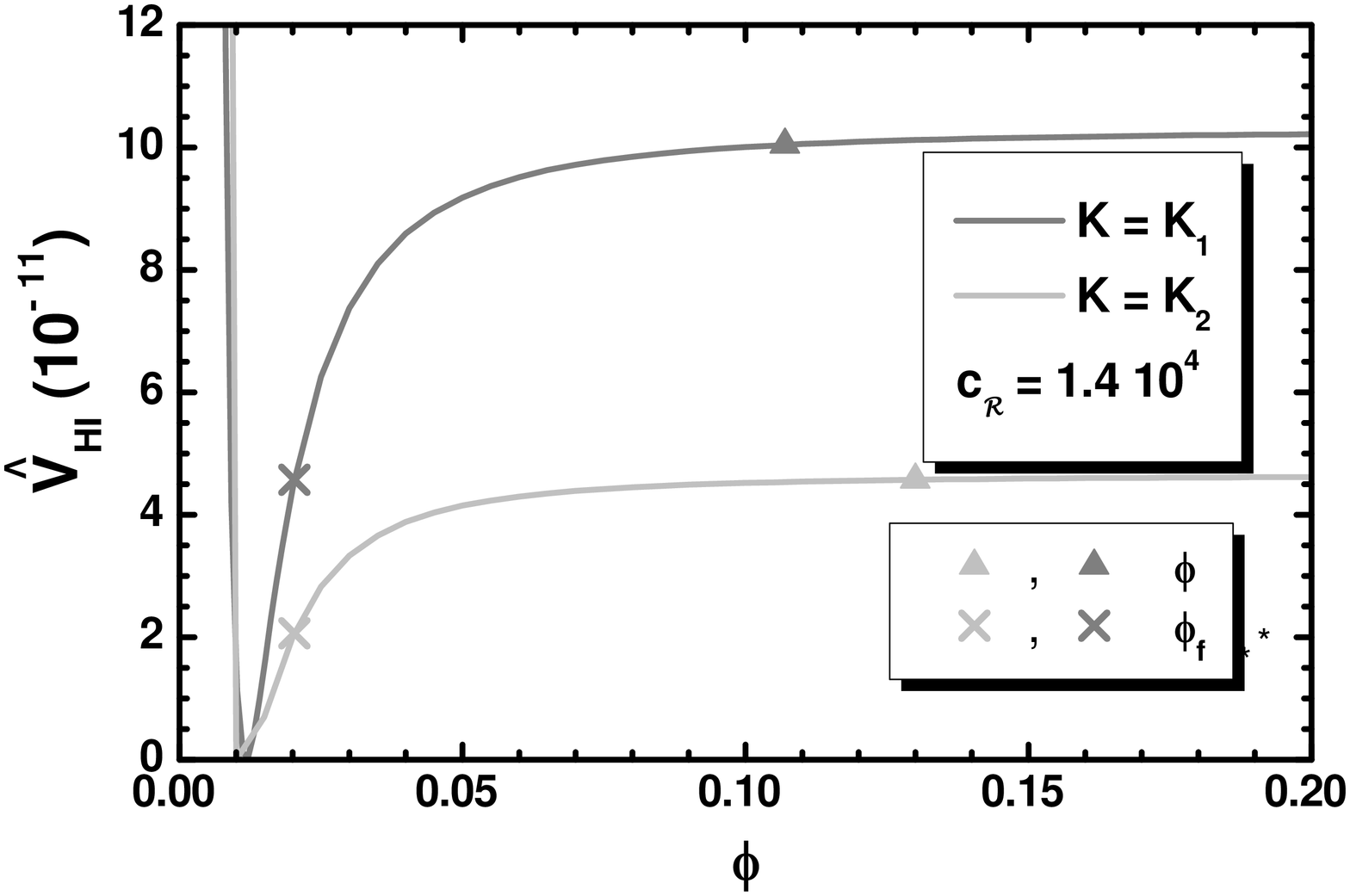,height=3.65in,angle=-90}\eec\vspace*{-.12in}
\hfill \caption{\sl\small Inflationary potential $\Vhi$ as a
function of $\sg$ for $\sg>0$, $\ck$ in Eq. (4.10) and $K=\kar$
(dark gray line) or $K=\kbr$ (light gray line) -- the values of
$\sgx$ and $\sgf$ are also indicated.}\label{fig1}
\end{figure}

The normalization of the amplitude, $\As$, of the power spectrum
of the curvature perturbations generated by $\sg$ at the pivot
scale $\ks$ allows us to determine $\ld$ as follows
\beq \label{Proba}\sqrt{\As}=\: \frac{1}{2\sqrt{3}\, \pi} \;
\frac{\Ve_{\rm IHI}(\sex)^{3/2}}{|\Ve_{\rm
IHI,\se}(\sex)|}=4.58\cdot10^{-5}\>\>\Rightarrow\>\> \ld
=32\pi\sqrt{6N\As}
\ck\frac{\Ns}{(4\Ns-N)^{2}}\simeq\begin{cases}0.29,&K=K_1,\\0.24,&K=K_2.\end{cases}
\eeq
The resulting relation reveals that $\ld$ is proportional to
$\ck$. For these $\ld$ values we display $\Vhi$ as a function of
$\sg$ in \Fref{fig1}. We observe that $\Vhi$ is a monotonically
increasing function of $\sg$. The inflationary scale,
$\Vhi^{1/4}$, approaches the SUSY GUT scale in \Eref{Mgut1} and
lies well below $\Qef=1$, consistently with the classical
approximation to the inflationary dynamics.

%%%%%%%%%%%%%%%%%%%%%%%%%%%%%%%%%%%%%%%%%%%%%%%%%%%%%%%%%%%%%%%%%%%%%
\renewcommand{\arraystretch}{1.}

%%%%%%%%%%%%%%%%%%%%%%%%%%%%%%%%%%%%%%%%%%

At the pivot scale, we can also calculate the scalar spectral
index, $\ns$, its running, $\as$, and the tensor-to-scalar ratio,
$r$, via the relations
\beqs\baq \label{ns} && \ns=\: 1-6\widehat\epsilon_\star\ +\
2\widehat\eta_\star
\simeq1-\frac{2}{\Ns}=0.963,~~~r=16\widehat\epsilon_\star\simeq
\frac{4N}{\Ns^2}=0.0032~[0.0022],\\
&& \label{as} \as =\:{2\over3}\left(4\widehat\eta_\star^2-(n_{\rm
s}-1)^2\right)-2\widehat\xi_\star\simeq-\frac{2}{\Ns^{2}}-\frac{7N}{2\Ns^3}=-0.005
~~~\mbox{for}~~~K=K_1~~[K_2]\eaq\eeqs
with $\widehat\xi={\Ve_{\rm IHI,\widehat\sg} \Ve_{\rm
IHI,\widehat\sg\widehat\sg\widehat\sg}/\Ve_{\rm IHI}^2}$ and the
variables with subscript $\star$ are being evaluated at
$\sg=\sgx$. The numerical values are obtained employing
$\Ns\simeq(57.5-60)$ which corresponds to a quartic potential. It
is expected to approximate $\Vhi$ rather well for $\sg\ll1$
\cite{ighi}.

The results above turn out to be in nice agreement with the
fitting of the \plk\ (release 4) \cite{plin}, baryon acoustic
oscillations, cosmic microwave background lensing and \bcp\ data
\cite{gwsnew} with the $\Lambda$CDM$+r$ model, i.e.,
\begin{equation}  \label{nswmap}
\mbox{\ftn\sf
(a)}\>\>\ns=0.965\pm0.009\>\>\>~\mbox{and}\>\>\>~\mbox{\ftn\sf
(b)}\>\>r\leq0.032,
\end{equation}
at 95$\%$ \emph{confidence level} ({\sf\ftn c.l.}) with
$|\as|\ll0.01$.

\section{IHI and $\mu$ Term of MSSM}
\label{secmu}

A byproduct of our setting is that it assists us to understand the
origin of $\mu$ term of MSSM, as we show in \Sref{secmu1},
consistently with the low-energy phenomenology of MSSM  -- see
\Sref{pheno}. Hereafter we restore units, i.e., we take
$\mP=2.433\cdot10^{18}~\GeV$.

\subsection{Generation of the $\mu$ Term of MSSM}\label{secmu1}

The contributions from the soft SUSY breaking terms, although
negligible during IHI, since these are much smaller than
$\sg\sim\mP$, may shift slightly $\vev{S}$ from zero in
\Eref{vevs}. Indeed, the relevant potential terms are
\beq V_{\rm soft}= \lf\ld A_\ld S \phcb\phc+\lm A_\mu S \hu\hd +
\ld_{iN^c} A_{iN^c}\phc \widetilde N^{c2}_i- {\rm a}_{S}S\ld M^2/4
+ {\rm h. c.}\rg+ m_{\gamma}^2\left|X^\gamma\right|^2,
\label{Vsoft} \eeq
where $m_{\gamma}, A_\ld, A_\mu, A_{iN^c}$ and $\aS$ are soft SUSY
breaking mass parameters.  Rotating $S$ in the real axis by an
appropriate $R$-transformation, choosing conveniently the phases
of $\Ald$ and $\aS$ so as the total low energy potential $V_{\rm
tot}=V_{\rm SUSY}+V_{\rm soft}$ to be minimized -- see \Eref{VF}
-- and substituting in $V_{\rm soft}$ the $\phc$  and $\phcb$
values from \Eref{vevs} we get
\beqs\beq \vev{V_{\rm tot}(S)}=
\ld^2\frac{(N\ck-1)M^4S^2}{4N^2\mP^2\ck} -\ld M^2S\am \mgr
~~\mbox{with}~~~\am=\lf|A_\ld| + |{\rm a}_{S}|\rg/2\mgr,
\label{Vol} \eeq
where $\mgr$ is the gravitino ($\Gr$) mass and $\am>0$ is a
parameter of order unity which parameterizes our ignorance for the
dependence of $|A_\ld|$ and $|{\rm a}_{S}|$ on $\mgr$. We also
take into account that $m_S\ll M$.  The extermination condition
for $\vev{V_{\rm tot}(S)}$ w.r.t $S$ leads to a non vanishing
$\vev{S}$ as follows
\beq \label{vevS}{d}\vev{V_{\rm tot}(S)}/{d S}
=0~~~\Rightarrow~~~\vev{S}\simeq {N\ck}\am\mgr/{\ld},\eeq\eeqs
where we employed \Eref{ig}. The extremum above is a global
minimum since ${d^2} \vev{V_{\rm tot}(S)}/{d
S^2}=2\ld^2\mP^2/\ck(N\ck-1)>0$. The generated $\mu$ term from the
term in \Eref{Wmu} is \beq\mu =\lm \vev{S} \simeq
\frac{\lm}{32\pi}\sqrt{\frac{N}{6\As}}
\frac{(4\Ns-N)^{2}}{\Ns}\am\mgr,\label{mu}\eeq
where we make use of \Eref{Proba} which reveals that the resulting
$\mu$ above does not depend on $\ld$ and $\ck$. Thanks to the
presence of $\sqrt{\As}\sim10^{-5}$ in the denominator any
$\mu/\mgr<1$ value is accessible for $\lm<10^{-5}$ which is
allowed by \Eref{lmb} without causing any ugly hierarchy between
$\mgr$ and $\mu$. On the other hand, given that $\mgr$ is
currently constrained beyond the $\TeV$ region a mild hierarchy
between $\mu$ and $\mgr$ assists us to alleviate the little
hierarchy problem ameliorating the naturalness of SUSY models
after the LHC Higgs discovery \cite{baer}.

\subsection{Connection with the MSSM Phenomenology}\label{pheno}

The SUSY breaking effects, considered in \Eref{Vsoft}, explicitly
break $U(1)_R$ to a subgroup, $\mathbb{Z}_2^{R}$ which remains
unbroken by $\vev{S}$ in \Eref{vevS} and so no disastrous domain
walls are formed. Combining $\mathbb{Z}_2^{R}$ with the
$\mathbb{Z}_2^{\rm f}$ fermion parity, under which all fermions
change sign, yields the well-known $R$-parity. This residual
symmetry prevents rapid proton decay and guarantees the stability
of the \emph{lightest SUSY particle} ({\sf\ftn LSP}), providing
thereby a well-motivated \emph{cold dark matter} ({\sf\ftn CDM})
candidate.

The candidacy of LSP may be successful, if its abundance is
consistent with the expectations for it from the $\Lambda$CDM
model \cite{plcp} within a concrete low energy framework. We here
adopt the \emph{Constrained MSSM} ({\ftn\sf CMSSM}), which is
relied on the following free parameters
\begin{equation}
{\rm
sign}\mu,~~\tan\beta=\vev{\hu}/\vev{\hd},~~\mg,~~m_0~~\mbox{and}~~A_0,
\label{para}
\end{equation}
where ${\rm sign}\mu$ is the sign of $\mu$, and the three last
mass parameters denote the common gaugino mass, scalar mass and
trilinear coupling constant, respectively, defined (normally) at
$\Mgut$. Imposing a number of cosmo-phenomenological constraints
-- from which the consistency of LSP relic density with
observations plays a central role -- the best-fit values of
$|A_0|$, $m_0$ and $|\mu|$ can be determined as in \cref{mssm}.
Their results are listed in the first four lines of \Tref{tab}. We
see that there are four allowed regions characterized by the
specific mechanism for suppressing the relic density of the LSP
which is the lightest neutralino ($\chi$) -- $\tilde\tau_1, \tilde
t_1$ and $\tilde \chi^\pm_1$ stand for the lightest stau, stop and
chargino eigenstate whereas $A$ and $H$ are the CP-odd and the
heavier CP-even Higgs bosons of MSSM respectively. The proposed
regions pass all the currently available LHC bounds \cite{lhc} on
the masses of the various sparticles.

%In regions (I), (II) and (IV) it is obtained $m_{\tilde
%g}\geq2.9~\TeV$, $m_{\tilde \chi^\pm}\geq1.1~\TeV$ \& $m_{\tilde
%t_1}\geq3.6~\TeV$ and so, these are still allowed by all the
%relevant requirements -- region (III) requires lower $m_{\tilde
%t_1}\simeq0.59~\TeV$ and so it is less favored.

Enforcing the conditions for the electroweak symmetry breaking a
value for the parameter $|\mu|$ can be achieved in each of the
regions in \Tref{tab}. Taking this $|\mu|$ value as input we can
extract the $\lm$ values, if we first derive $\am$ setting, e.g.,
\beq
m_0=\mgr~~~\mbox{and}~~~|A_0|=|A_\ld|=|\aS|.\label{softass}\eeq
Here we ignore possible renormalization group effects. The outputs
of our computation is listed in the two rightmost columns of
\Tref{tab} for $K=K_1$ and $K_2$. From these we infer that the
required $\lm$ values, in all cases besides the one, written in
italics, are comfortably compatible with \Eref{lmb} for $\nsu=2$
which imply $\lm\lesssim2\cdot10^{-5}$. Concluding, the whole
inflationary scenario can be successfully combined with all the
allowed regions CMSSM besides region (II) for $K=K_1$. On the
other hand, regions (I) \& (IV) are more favored from the point of
view of the $\Gr$ constraint. Indeed, only for $\mgr\gtrsim9~\TeV$
the unstable $\Gr$  becomes cosmologically safe for the $\Trh$
values, necessitated for satisfactory nTL -- see \eqs{Ygw}{res2}
in \Sref{num} below.

\renewcommand{\arraystretch}{1.25}
\begin{table}[!t] \bec
\begin{tabular}{|lc|c|c|c||c|c|c|}\hline
\multicolumn{2}{|c|}{\sc CMSSM
}&$|A_0|$&$m_0$&$|\mu|$&$\am$&\multicolumn{2}{|c|}{
$\lm$~({\boldmath $10^{-6}$})}\\\cline{7-8}
\multicolumn{2}{|c|}{Region}&$(\TeV)$&$(\TeV)$&$(\TeV)$&&$K=K_1$&$K=K_2$
\\\hline\hline
{\bfseries(I)}&$A/H$ Funnel &$9.9244$ &$9.136$&$1.409$&$1.086$ &{\boldmath $0.963$}&{\boldmath
$1.184$}\\
{\bfseries(II)}&$\tilde\tau_1-\chi$ Coannihilation &$1.2271$ &$1.476$&$2.62$& $0.831$&{\boldmath ${\it 14.48}$}&{\boldmath $17.81$ }\\
{\bfseries(III)}&$\tilde t_1-\chi$ Coannihilation  &$9.965$ &$4.269$&$4.073$&$2.33$ &{\boldmath $2.91$}&{\boldmath $3.41$}\\
{\bfseries(IV)}&$\tilde \chi^\pm_1-\chi$ Coannihilation  &$9.2061$ &$9.000$&$0.983$&$1.023$ &{\boldmath $0.723$}&{\boldmath $0.89$}\\
\hline
\end{tabular}
\end{center}
\caption[]{\sl\small The required $\lm$ values which render our
models compatible with the best-fit points in the CMSSM, as found
in \cref{mssm}, for the assumptions of \Eref{softass}, $\nsu=2$,
and $K=K_1$ or $K=K_2$.} \label{tab}
\end{table}\renewcommand{\arraystretch}{1.}

\section{Non-Thermal Leptogenesis and Neutrino Masses}\label{pfhi}

We below specify how our inflationary scenario makes a transition
to the radiation dominated era (\Sref{lept0}) and offers an
explanation of the observed BAU (\Sref{lept1}) consistently with
the $\Gr$ constraint and the low energy neutrino data. Our results
are summarized in \Sref{num}.

\subsection{Inflaton Mass \& Decay}\label{lept0}

Soon after the end of IHI, the (canonically normalized) inflaton
\beq\dphi=\vev{J}\dph\>\>\>\mbox{with}\>\>\> \dph=\phi-M
\>\>\>\mbox{and}\>\>\>\vev{J}=\sqrt{N\ck}\label{dphi} \eeq
acquires mass given by
\beq \label{msn} %
\msn=\left\langle\Ve_{\rm IHI,\se\se}\right\rangle^{1/2}=
\left\langle \Ve_{\rm
IHI,\sg\sg}/J^2\right\rangle^{1/2}\simeq\frac{\ld
\mP}{\sqrt{\ck\lf{N\ck}-1\rg}}\simeq2.8\cdot10^{4}~\EeV,\eeq
where $1~\EeV=10^9~\GeV$.  This value is equal to that encountered
in other models of induced-gravity inflation \cite{R2r, nIG} and
larger than those obtained in several versions of non-minimal
\cite{univ} or pole-induced \cite{so} Higgs inflation. Also
$\dphi$ settles into a phase of damped oscillations abound the
minimum in \Eref{vevs} reheating the universe at a temperature
\cite{ighi}
\beq\Trh=
\left({72/5\pi^2g_{*}}\right)^{1/4}\lf\Gsn\mP\rg^{1/2}\>\>\>\mbox{with}\>\>\>\Gsn=\GNsn+\Ghsn+\Gysn\,.\label{Trh}\eeq
Also $g_{*}=228.75$ counts the MSSM effective number of
relativistic degrees of freedom and we take into account the
following decay widths
\beqs\bea \GNsn&=&\frac{g_{iN^c}^2}{16\pi}\msn\lf1-\frac{4\mrh[
i]^2}{\msn^2}\rg^{3/2}\>\>\mbox{with}\>\>\>
g_{iN^c}=(N-1)\frac{\ld_{iN^c}}{\vev{J}},\\
\Ghsn&=&\frac{2}{8\pi}g_{H}^2\msn\>\>\>\>\mbox{with}\>\>\>\>
g_{H}=\frac{\lm}{\sqrt{2}}, \\
\Gysn&=&\frac{14
g_y^2}{512\pi^3}\frac{\msn^3}{\mP^2}\>\>\>\>\mbox{with}\>\>\>\>g_y=y_{3}\lf\frac{N\ck-1}{2\ck}\rg^{1/2}\eea\eeqs
and $y_{3}=h_{t,b,\tau}(\msn)\simeq0.5$. Here $h_t, h_b$ and
$h_\tau$ are the Yukawa coupling constants $h_{3U}$, $h_{2D}$ and
$h_{3E}$ in \Eref{wmssm} respectively -- we assume that
diagonalization has been performed in the generation space. They
arise from the lagrangian terms
\beqs\bea {\cal L}_{\dphi\to \sni\sni}&=&
-\frac12e^{K/2\mP^2}W_{{\rm RHN},N_i^cN^c_i}\sni\sni\ +\ {\rm
h.c.}=g_{iN^c} \dphi\ \lf\sni\sni\ +\ {\rm h.c.}\rg +\cdots,\\
{\cal
L}_{\dphi\to\hu\hd}&=&-e^{K/\mP^2}K^{SS^*}\left|W_{\mu,S}\right|^2
=-g_{H} \msn\dphi \lf H_u^*H_d^*\ +\ {\rm h.c.}\rg+\cdots, \\
{\cal L}_{\dphi\to XYZ}&=&-g_y(\dphi/\mP)\lf
X\psi_{Y}\psi_{Z}+Y\psi_{X}\psi_{Z}+ Z\psi_{X}\psi_{Y}\rg+{\rm
h.c.}, \label{lint} \eea\eeqs
describing $\dphi$ decay into a pair of $N^c_{j}$ with masses
$\mrh[j]=\ld_{jN^c}M$, $\hu$ and $\hd$ and three MSSM
(s)-particles $X, Y, Z$, respectively.

\subsection{Lepton-Number and Gravitino Abundances}\label{lept1}

For $\Trh<\mrh[i]$, the out-of-equilibrium decay of $N^c_{i}$
generates a lepton-number asymmetry (per $N^c_{i}$ decay),
$\ve_i$. The resulting lepton-number asymmetry is partially
converted through sphaleron effects into a yield of the observed
BAU
\beq Y_B=-0.35\cdot{5\over2}{\Trh\over\msn}\mbox{$\sum_i$}
{\GNsn\over\Gsn}\ve_i~~~\mbox{with}~~~\ve_i =\sum_{j\neq i}\frac{
\im\left[(\mD[]^\dag\mD[])_{ij}^2\right]}{8\pi\vev{\hu}^2(\mD[]^\dag\mD[])_{ii}}\bigg(
F_{\rm S}\lf x_{ij},y_i,y_j\rg+F_{\rm
V}(x_{ij})\bigg).\label{Yb}\eeq
Here $\vev{\hu}\simeq174~\GeV$, for large $\tan\beta$, $F_{\rm S}$
[$F_{\rm V}$] are the functions entered in the vertex and
self-energy contributions computed as indicated in \cref{plum} and
$m_{\rm D}$ is the Dirac mass matrix of neutrinos, $\nu_i$,
arising from the forth term in \Eref{wmssm}. Employing the seesaw
formula we can then obtain the light-neutrino masses $m_{i\nu}$ in
terms of $\mD[i]$ and $\mrh[i]$ given by \Eref{Wrhn}. As a
consequence, nTL can be nicely linked to low energy neutrino data.
We take into account the recently updated best-fit values
\cite{valle} of that data listed in \Tref{tabn}. Furthermore, the
sum of $\mn[i]$'s is bounded from above at 95\% c.l. by the data
\cite{plcp, valle}
\beq\mbox{$\sum_i$} \mn[i]\leq0.23~{\eV}~~~\mbox{for NO
$\mn[i]$'s}~~~\mbox{or}~~~\mbox{$\sum_i$}
\mn[i]\leq0.15~{\eV}~~~\mbox{for IO $\mn[i]$'s},\label{sumnu}\eeq
where {\ftn\sf NO [IO]} stands for \emph{normal [inverted]
ordered} neutrino masses $\mn[i]$'s.

\renewcommand{\arraystretch}{1.1}
\begin{table}[!t]
\begin{center}
\begin{tabular}{|c|c|c|}\hline
{\sc Parameter }&\multicolumn{2}{c|}{\sc Best Fit
$\pm1\sigma$}\\\cline{2-3} &{\sc Normal}&{\sc Inverted}
\\\cline{2-3} &\multicolumn{2}{c|}{\sc Hierarchy} \\
\hline\hline
$\Delta
m^2_{21}/10^{-5}\eV^2~~~~$&\multicolumn{2}{c|}{$7.5^{+0.22}_{-0.20}$}\\\cline{2-3}
$\Delta
m^2_{31}/10^{-3}\eV^2$&$2.55^{+0.02}_{-0.03}$&{$2.45^{+0.02}_{-0.03}$}\\\hline
$\sin^2\theta_{12}/0.1$ &
\multicolumn{2}{c|}{$3.18\pm0.16$}\\\cline{2-3}
$\sin^2\theta_{13}/0.01$&$2.2^{+0.069}_{-0.062}$&$2.225^{+0.064}_{-0.070}$\\
$\sin^2\theta_{23}/0.1$&$5.74\pm0.14$&$5.78^{+0.10}_{-0.17}$\\\hline
$\delta/\pi~~$&$~~1.08^{+0.13}_{-0.12}~~~$&$~~~1.58^{+0.15}_{-0.16}~~~$\\
\hline
\end{tabular}\end{center}
\caption{\sl\small Low energy experimental neutrino data for
normal or inverted hierarchical neutrino masses. }\label{tabn}
\end{table}
\renewcommand{\arraystretch}{1.}

The validity of \Eref{Yb} requires that the $\dphi$ decay into a
pair of $\sni$'s is kinematically allowed for at least one species
of the $\sni$'s and also that there is no erasure of the produced
$Y_L$ due to $N^c_1$ mediated inverse decays and $\Delta L=1$
scatterings. These prerequisites are ensured if we impose
\beq\label{kin} {\sf \ftn
(a)}\>\>\msn\geq2\mrh[1]\>\>\>\mbox{and}\>\>\>{\sf \ftn
(b)}\>\>\mrh[1]\gtrsim 10\Trh.\eeq
Finally, \Eref{Yb} has to reproduce the observational result
\cite{plcp}
\beq Y_B=\lf8.697\pm0.054\rg\cdot10^{-11}.\label{BAUwmap}\eeq
The required $\Trh$ in \Eref{Yb} must be compatible with
constraints on the  $\Gr$ abundance, $Y_{3/2}$, at the onset of
\emph{nucleosynthesis} (BBN), which is estimated to be
\beq\label{Ygr} Y_{3/2}\simeq 1.9\cdot10^{-22}\ \Trh/\GeV,\eeq
where we take into account only thermal production of $\Gr$, and
assume that $\Gr$ is much heavier than the MSSM gauginos. On the
other hand, $\Yg$  is bounded from above in order to avoid
spoiling the success of the BBN. For the typical case where $\Gr$
decays with a tiny hadronic branching ratio, we have
\beq  \label{Ygw} \Yg\lesssim\left\{\bem
%\begin{array}{rl}
%
10^{-13}\hfill \cr
10^{-12}\hfill \cr \eem
%\end{array}
\right.\>\>\>\>\mbox{for}\>\>\>\>\mgr\simeq\left\{\bem
10.6~\TeV\hfill \cr
13.5~\TeV\hfill \cr \eem
%\end{array}
\right.\>\>\>\>\mbox{implying}\>\>\>\>\Trh\lesssim0.53\cdot\left\{\bem
%\begin{array}{rl}
%
1~\EeV\,,\hfill \cr
10~\EeV\,.\hfill \cr\eem
%\end{array}
\right.\eeq
The bounds above can be somehow relaxed in the case of a stable
$\Gr$.

\subsection{Results}\label{num}

Confronting $Y_B$ and $\Yg$ -- see \eqs{Yb}{Ygr} -- with
observations we can constrain the parameters of neutrino sector.
This is because $Y_B$ and $\Yg$ depend on $\msn$, $\Trh$,
$\mrh[i]$ and $\mD[i]$ and can interconnect IHI with neutrino
physics. We follow the bottom-up approach detailed in \cref{ighi},
according to which we find the $\mrh[i]$'s by using as inputs the
$\mD[i]$'s, a reference mass of the $\nu_i$'s -- $\mn[1]$ for NO
$\mn[i]$'s, or $\mn[3]$ for IO $\mn[i]$'s --, the two Majorana
phases $\varphi_1$ and $\varphi_2$ of the PMNS matrix, and the
best-fit values for the low energy parameters of neutrino physics
shown in \Tref{tabn}.

%On the other hand, running of the inflationary from the scale
%$\Lambda$ until $\Lambda_L$ is ignored.

%%%%%%%%%%%%%%%%%%%%%%%%%%%%%%%%%%%%%%%%%%%%%%%%%%%%%%%%%%%%%%%%%%%%%
\begin{figure}[!t]\vspace*{-.12in}
\hspace*{-.19in}
\begin{minipage}{8in}
\epsfig{file=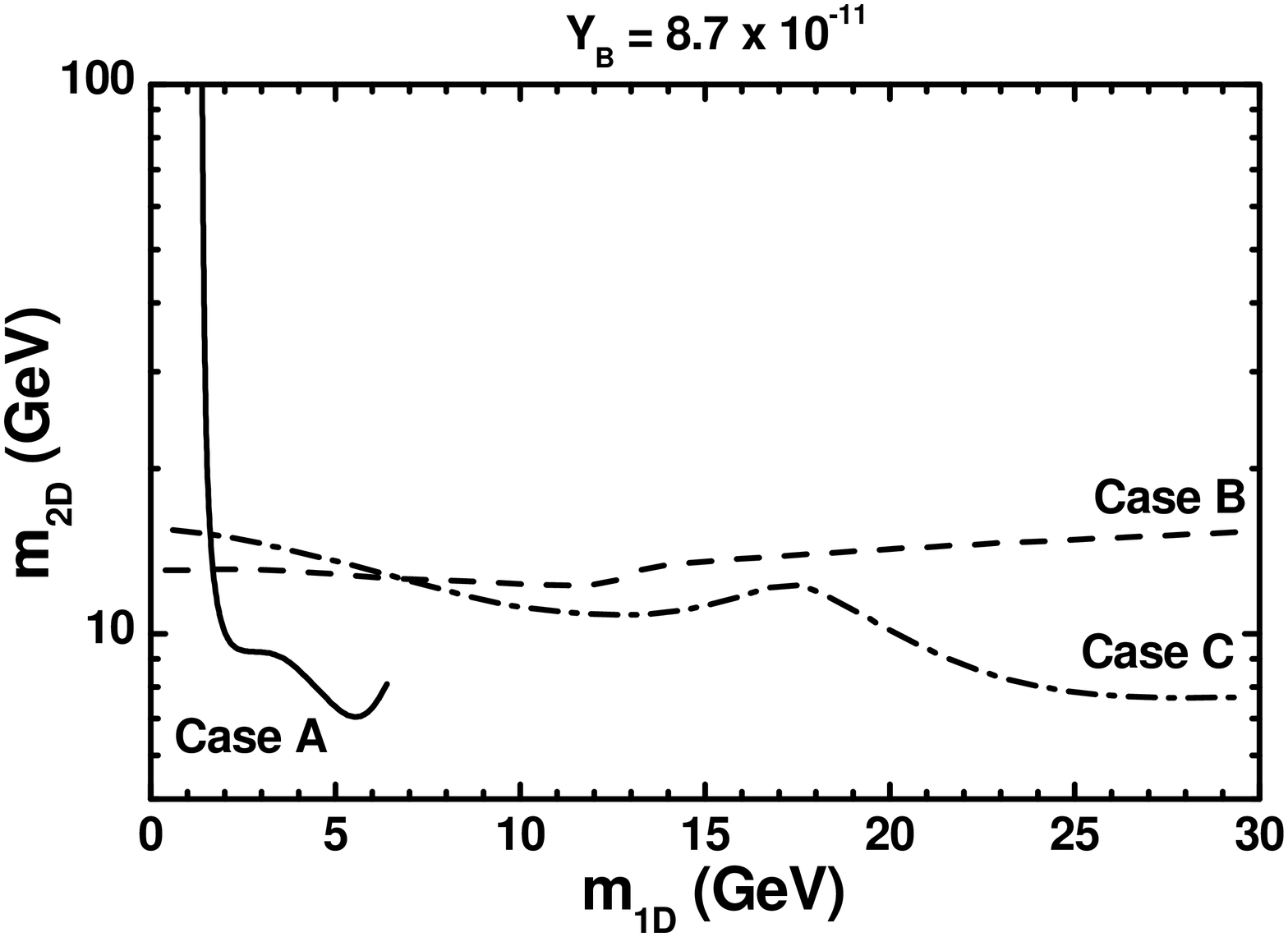,height=3.5in,angle=-90}
\hspace*{-1.2cm}
\epsfig{file=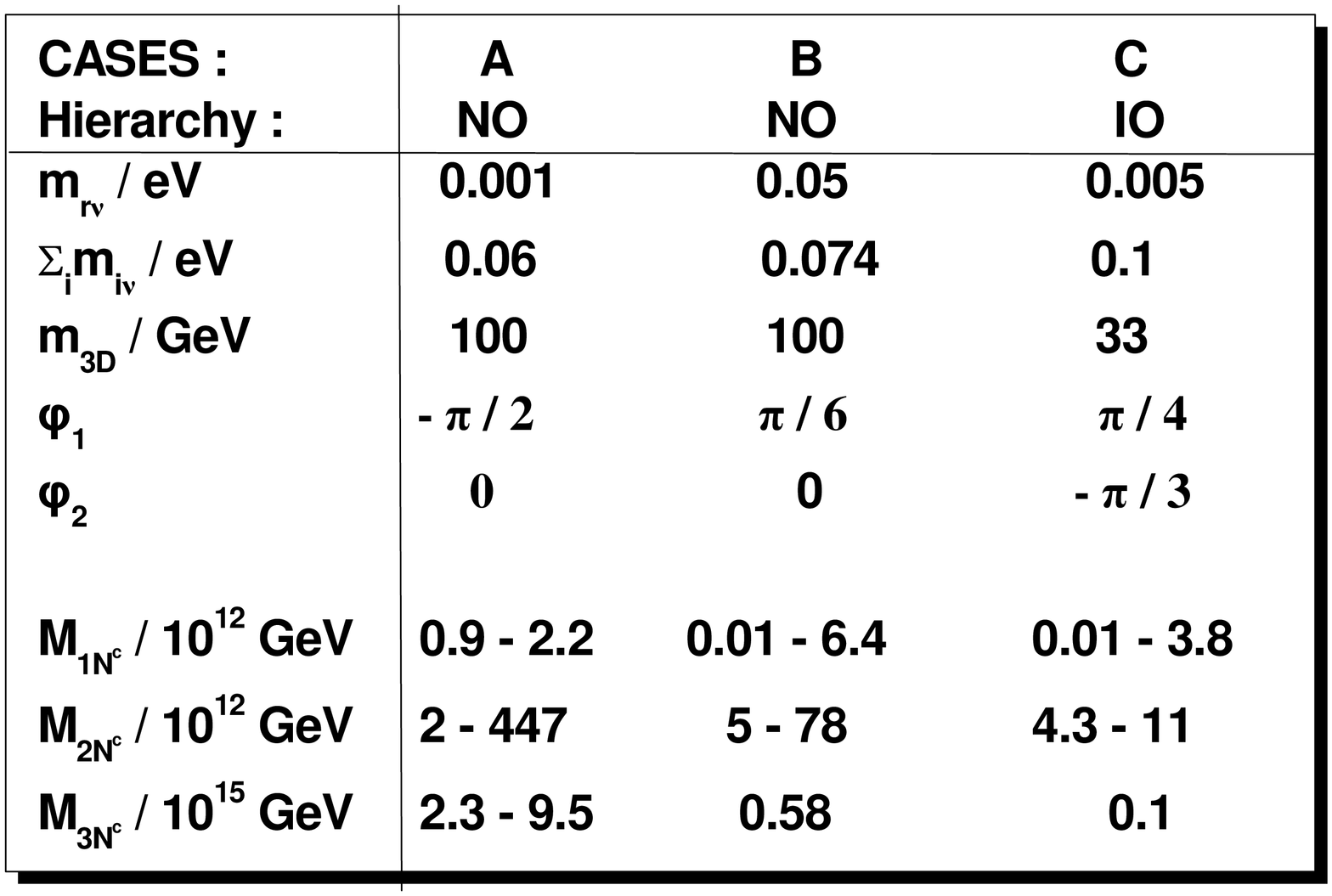,height=3.5in,angle=-90} \hfill
\end{minipage}
\hfill \caption{\sl\small  Contours, yielding the central $Y_B$ in
Eq.~(6.9) consistently with the inflationary requirements, in the
$m_{\rm 1D}-m_{\rm 2D}$ plane. We take $K=K_2$ with $\nsu=2$,
$\lm=10^{-6}$ and the values of $m_{i\nu}$, $m_{\rm 1D}$, $m_{\rm
3D}$, $\varphi_1$ and $\varphi_2$ which correspond to the cases A
(solid line), B (dashed line) and C (dot-dashed line).}\label{fmD}
\end{figure}
%%%%%%%%%%%%%%%%%%%%%%%%%%%%%%%%%%%%%%%%%%%%%%%%%%%%%%%%%%%%%%%%%%%%%

The outcome of our computation is presented in \Fref{fmD}, where
we depict the allowed values of $m_{\rm 2D}$ versus $m_{\rm 1D}$
for $K=K_2$ with $\nsu=2$, $\lm=10^{-6}$ and the remaining
parameters shown in the Table of \Fref{fmD}. The conventions
adopted for the various lines is depicted in the plot label. In
particular, we use solid, dashed and dot-dashed line when the
remaining inputs -- i.e. $\mn[i]$, $\mD[3]$, $\varphi_1$, and
$\varphi_2$ -- correspond to the cases A, B and C of the Table of
\Fref{fmD} respectively. We consider NO (cases A and B) and IO
(case C) $\mn[i]$'s. In all cases, the current limit in
\Eref{sumnu} is safely met. The gauge symmetry considered here
does not predict any particular Yukawa unification pattern and so,
the $\mD[i]$'s are free parameters. This fact offers us a
convenient flexibility for the fulfilment of all the imposed
requirements. Care is also taken so that the perturbativity of
$\ld_{iN^c}$ holds, i.e., $\ld_{iN^c}^2/4\pi\leq1$. The inflaton
$\dphi$ decays mostly into $N_1^c$'s. In all cases
$\GNsn<\Ghsn<\Gysn$ and so the ratios $\GNsn/\Gsn$ in \Eref{Yb}
introduce a considerable reduction in the derivation of $\Yb$. For
the considered cases in \Fref{fmD} we obtain:
\beqs\beq  0.01\lesssim
M_{1N^c}/10^{3}~\EeV\lesssim6.4,\>\>\>2\lesssim
M_{2N^c}/10^{3}~\EeV\lesssim447\>\>\>\mbox{and}\>\>\>0.1\lesssim
M_{2N^c}/10^{6}~\EeV\lesssim9.5.\label{res1}\eeq
As regards the other quantities, in all we obtain
\beq  1.4\lesssim
Y_{\Gr}/10^{-13}\lesssim1.7\>\>\>\mbox{with}\>\>\>0.75\lesssim\Trh/{\EeV}\lesssim0.9\,.\label{res2}\eeq\eeqs
As a bottom line, nTL is a realistic possibility within our
setting provided that $\mgr\sim10~\TeV$ as deduced from
\eqs{Ygw}{res2}. As advertised in \Sref{pheno}, these values are
in nice agreement with the ones needed for the solution of the
$\mu$ problem within CMSSM in regions (I) and (IV) of \Tref{tab}.

\section{Conclusions}\label{con}

We investigated the realization of IHI in the framework of a $B-L$
extension of MSSM endowed with the condition that the GUT scale is
determined by the renormalization-group running of the three gauge
coupling constants. Our setup is tied to the super-{} and \Kap s
given in Eqs.~(\ref{Wtot}) and (\ref{K1}) -- (\ref{K2}).  Our
models exhibit the following features:

\begin{itemize}

\item[{\sf\ftn (i)}] they predict the correct $\ns$ and low $r$
thanks to the induced-gravity and the GUT requirements;

\item[{\sf\ftn (ii)}] they ensure the validity of the effective
theory up-to $\mP$;

\item[{\sf\ftn (iii)}] they inflate away cosmological defects;

\item[{\sf\ftn (iv)}] they offer a nice solution to the $\mu$
problem of MSSM, provided that $\ld_\mu$ is somehow small;

\item[{\sf\ftn (v)}] they allow for baryogenesis via nTL
compatible with $\Gr$ constraints and neutrino data. In
particular, we may have $\mgr\sim10~\TeV$, with the inflaton
decaying mainly to $N^c_1$ and $N^c_2$ -- we obtain $\mrh[i]$ in
the range $(10^{10}-10^{15})~\GeV$.

\end{itemize}

It remains to introduce a consistent soft SUSY breaking sector --
see, e.g., \cref{susyb} -- to obtain a self-contained theory --
cf. \cref{nsreview, lhclinde}. Moreover, since our main aim here
is the demonstration of the mechanism of IHI in SUGRA, we opted to
utilize the simplest GUT embedding. Extensions to more structured
GUTs are also possible -- see e.g. \cref{igsu5, nmh} -- with
similar inflationary observables.

\newpage

\paragraph*{\small \bf\scshape Acknowledgments} {\small I would like to thank H. Baer and S. Ketov
and for interesting discussions. This research work was supported
by the Hellenic Foundation for Research and Innovation (H.F.R.I.)
under the ``First Call for H.F.R.I. Research Projects to support
Faculty members and Researchers and the procurement of high-cost
research equipment grant'' (Project Number: 2251).}

\def\ijmp#1#2#3{{\emph{Int. Jour. Mod. Phys.}}
{\bf #1},~#3~(#2)}
\def\plb#1#2#3{{\emph{Phys. Lett.  B }}{\bf #1},~#3~(#2)}
\def\zpc#1#2#3{{Z. Phys. C }{\bf #1},~#3~(#2)}
\def\prl#1#2#3{{\emph{Phys. Rev. Lett.} }
{\bf #1},~#3~(#2)}
\def\rmp#1#2#3{{Rev. Mod. Phys.}
{\bf #1},~#3~(#2)}
\def\prep#1#2#3{\emph{Phys. Rep. }{\bf #1},~#3~(#2)}
\def\prd#1#2#3{{\emph{Phys. Rev.  D }}{\bf #1},~#3~(#2)}
\def\npb#1#2#3{{\emph{Nucl. Phys.} }{\bf B#1},~#3~(#2)}
\def\npps#1#2#3{{Nucl. Phys. B (Proc. Sup.)}
{\bf #1},~#3~(#2)}
\def\mpl#1#2#3{{Mod. Phys. Lett.}
{\bf #1},~#3~(#2)}
\def\arnps#1#2#3{{Annu. Rev. Nucl. Part. Sci.}
{\bf #1},~#3~(#2)}
\def\sjnp#1#2#3{{Sov. J. Nucl. Phys.}
{\bf #1},~#3~(#2)}
\def\jetp#1#2#3{{JETP Lett. }{\bf #1},~#3~(#2)}
\def\app#1#2#3{{Acta Phys. Polon.}
{\bf #1},~#3~(#2)}
\def\rnc#1#2#3{{Riv. Nuovo Cim.}
{\bf #1},~#3~(#2)}
\def\ap#1#2#3{{Ann. Phys. }{\bf #1},~#3~(#2)}
\def\ptp#1#2#3{{Prog. Theor. Phys.}
{\bf #1},~#3~(#2)}
\def\apjl#1#2#3{{Astrophys. J. Lett.}
{\bf #1},~#3~(#2)}
\def\n#1#2#3{{Nature }{\bf #1},~#3~(#2)}
\def\apj#1#2#3{{Astrophys. J.}
{\bf #1},~#3~(#2)}
\def\anj#1#2#3{{Astron. J. }{\bf #1},~#3~(#2)}
\def\mnras#1#2#3{{MNRAS }{\bf #1},~#3~(#2)}
\def\grg#1#2#3{{Gen. Rel. Grav.}
{\bf #1},~#3~(#2)}
\def\s#1#2#3{{Science }{\bf #1},~#3~(#2)}
\def\baas#1#2#3{{Bull. Am. Astron. Soc.}
{\bf #1},~#3~(#2)}
\def\ibid#1#2#3{{\it ibid. }{\bf #1},~#3~(#2)}
\def\cpc#1#2#3{{Comput. Phys. Commun.}
{\bf #1},~#3~(#2)}
\def\astp#1#2#3{{Astropart. Phys.}
{\bf #1},~#3~(#2)}
\def\epjc#1#2#3{{Eur. Phys. J. C}
{\bf #1},~#3~(#2)}
\def\nima#1#2#3{{Nucl. Instrum. Meth. A}
{\bf #1},~#3~(#2)}
\def\jhep#1#2#3{{\emph{J. High Energy Phys.} }
{\bf #1},~#3~(#2)}
\def\jcap#1#2#3{{\emph{J. Cosmol. Astropart. Phys.} }
{\bf #1},~#3~(#2)}
\def\jcapn#1#2#3#4{{\sl J. Cosmol. Astropart. Phys. }{\bf #1}, no. #4, #3 (#2)}
\def\prdn#1#2#3#4{{\sl Phys. Rev. D }{\bf #1}, no. #4, #3 (#2)}
\newcommand{\arxiv}[1]{{\ftn\tt  arXiv:#1}}
\newcommand{\hepph}[1]{{\ftn\tt  hep-ph/#1}}
\newcommand{\astroph}[1]{{\ftn\tt  astro-ph/#1}}
\def\prdn#1#2#3#4{{\sl Phys. Rev. D }{\bf #1}, no. #4, #3 (#2)}
\def\jcapn#1#2#3#4{{\sl J. Cosmol. Astropart.
Phys. }{\bf #1}, no. #4, #3 (#2)}
\def\epjcn#1#2#3#4{{\sl Eur. Phys. J. C }{\bf #1}, no. #4, #3 (#2)}

\end{document}